\pdfoutput=1
\documentclass[11pt,twoside,a4paper,cmspaper,final,collab]{cms-tdr}
\def\svnVersion{37c0bfe}\def\svnDate{2026/11/24}\def\cmsCernNoTag{CERN-EP-2026-169}\def\cmsCernDate{\today}\def\cmsMessage{Published in the Journal of High Energy Physics as \href{http://dx.doi.org/10.1007/JHEP11(2026)147}{\doi{10.1007/JHEP11(2026)147}.}}
\begin{document}\cmsNoteHeader{EXO-24-001}

\newcommand{\chione}{\HepParticle{\upchi}{1}{}\xspace}
\newcommand{\chitwo}{\HepParticle{\upchi}{2}{}\xspace}
\newcommand{\chitwochitwo}{\ensuremath{\chitwo\chitwo}\xspace}
\newcommand{\yzero}{\HepParticle{Y}{0}{}\xspace}
\newcommand{\yone}{\HepParticle{Y}{1}{}\xspace}
\newcommand{\coup}{\ensuremath{g_{\chitwo\chione\yzero}}\xspace}
\newcommand{\wj}{\ensuremath{\PW+\text{jets}}\xspace}
\newcommand{\zj}{\ensuremath{\PZ+\text{jets}}\xspace}
\newcommand{\zvvj}{\ensuremath{\PZ(\to\PGn\PGn)+\text{jets}}\xspace}
\newcommand{\sgnn}{\ensuremath{s_{\text{GNN}}}\xspace}
\newcommand{\MADSPIN}{\textsc{MadSpin}\xspace}
\newlength\cmsTabSkip\setlength{\cmsTabSkip}{1ex}

\cmsNoteHeader{EXO-24-001}

\title{Search for dark matter in a signature with a four-prong large-radius jet in proton-proton collisions at \texorpdfstring{$\sqrt{s}=13\TeV$}{sqrt(s)=13 TeV}}

\author*[cern]{A. Cern Person}

\date{\today}

\abstract{
A search for a pair of nonprompt dark matter (DM) candidates produced in association with an initial-state radiation jet, in a signature containing a four-prong large-radius jet, is presented. The signal model contains a heavy vector or axial-vector mediator, which produces long-lived dark-sector particles that decay to a stable DM particle and a light boson, which decays to quarks. The analysis is based on data collected in the years 2016--2018 with the CMS detector at the LHC in proton-proton collisions at $\sqrt{s}=13\TeV$, corresponding to an integrated luminosity of 138\fbinv. Signal candidates feature large-radius jets, which are identified using a jet substructure tagger based on a graph neural network. The large-radius jet aims to reconstruct the decay of light DM mediators into four quarks, which are produced in association with two stable DM particles. The standard model background contributions are estimated from data using dedicated control regions. The missing transverse momentum spectrum is probed for a potential signal over the expected background. No significant excess over the standard model expectation is observed. Upper limits at 95\% confidence level are set on the signal strength as functions of either the mediator mass or the relevant coupling. This is the first search for a pair of nonprompt DM candidates in the Lorentz-boosted topology, characterized by a large-radius jet and large missing transverse momentum.
}

\hypersetup{
pdfauthor={CMS Collaboration},
pdftitle={Search for dark matter produced in association with a large-radius jet in proton-proton collisions at sqrt{s} = 13 TeV},
pdfsubject={CMS},
pdfkeywords={CMS, BSM, dark matter}} 

\maketitle 

\section{Introduction}

\begin{figure}[!htbp]
  \centering
  \includegraphics[width=0.9\textwidth]{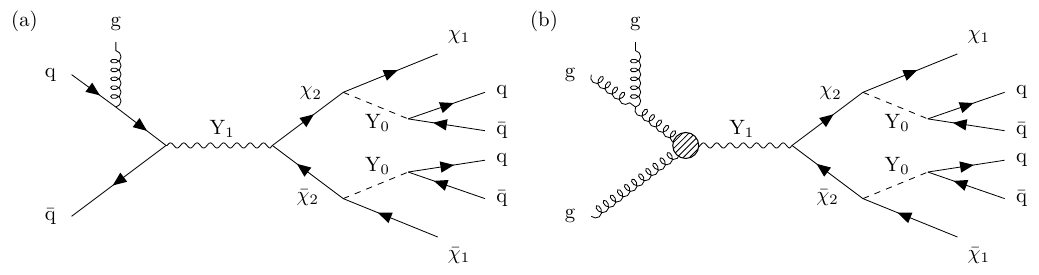}
  \caption{Representative Feynman diagrams of the signal process. The mediator \yone is produced via quark-antiquark annihilation (left) or gluon-gluon fusion (right) in association with a jet from initial-state radiation. A hatched circle represents an effective $\mathrm{Y}_1 \mathrm{g} \mathrm{g}$ coupling.}
    \label{fig:1-diagram1}
\end{figure}

While the nature of dark matter (DM) remains unknown, many extensions of the standard model (SM) predict a richer dark sector containing additional, potentially long-lived particles that can be produced at colliders and decay at macroscopic distances within the detectors. This paper investigates nonprompt DM signatures, which can be regarded as an extension of the simplified DM models~\cite{simpdm}, as outlined in Ref.~\cite{displaceddm}. Dark sectors need not consist only of a single stable particle, but may contain heavier neutral states whose transitions to the stable DM state are suppressed. Such transitions can result in macroscopic lifetimes, leading to displaced visible decays accompanied by missing transverse momentum ($\ptmiss$). Simplified models combining long-lived dark-sector states with $\ptmiss$ provide a common phenomenological description of signatures that can arise in complete scenarios such as gauge-mediated supersymmetry breaking and neutral-naturalness models, and which are not generally covered optimally by prompt-DM searches. The model used in this paper incorporates displaced secondary vertices characterized by the mass and lifetime of long-lived particles. Feynman diagrams of the studied process are shown in Fig.~\ref{fig:1-diagram1}. A pair of Dirac fermions \chitwo is produced via a vector or axial-vector mediator \yone. Each \chitwo particle decays into a stable fermion \chione and a scalar particle \yzero. The \yzero boson is assumed to subsequently decay into a pair of SM quarks. Depending on the choice of the model parameters, the \chitwo particle may be long-lived, resulting in a nonprompt decay with respect to the primary interaction point. In this model, DM production proceeds via a heavy mediator that couples the SM to a dark sector containing additional particles. The characteristic signature arises from the production of long-lived particles that decay nonpromptly within the detector  (``displaced decays"), giving rise to displaced hadronic activity accompanied by missing transverse momentum (\ptmiss) from the two undetected stable DM particles \chione.

In this search we investigate the topology in which the mediator \yone is produced in association with a jet from initial-state radiation (ISR) with high transverse momentum \pt. This requirement enables efficient selection of candidate signal events, despite the relatively high thresholds of jet-based triggers in the CMS experiment at the CERN LHC, and enables the exploration of regions with low mediator mass. In all benchmark scenarios considered in this analysis, the \yzero boson mass is fixed at 1\GeV. At this mass, the \yzero boson is assumed to decay promptly and predominantly into light-flavor quark pairs. Because of the large Lorentz boost of the \yone system and the small masses of the particles in the decay chain, the two \chitwo decays are expected to be collimated. The analysis therefore targets events in which the four quarks from the two \yzero decays are reconstructed as a single large-radius jet with a four-prong substructure.

For long \chitwo lifetimes, the decay products of the two \chitwo particles may originate from spatially separated displaced vertices; in this case, the large-radius jet reconstruction remains sensitive to the boosted topology, but the reconstruction of the jet substructure and subjet directions becomes less efficient. A jet tagger based on a graph neural network (GNN) is employed to identify signal-like jets based on their substructure. The tagger output is used to define the signal region (SR) and background-enriched control regions (CRs). The CRs are used to estimate the dominant background contributions from events composed uniquely of jets produced through the strong interaction, referred to as quantum chromodynamics (QCD) multijet events, as well as those from the production of a \PW or \PZ boson in association with jets (\wj and \zj, respectively).
Subdominant backgrounds, which include single and pair production of top quarks, and the production of two massive gauge bosons (diboson), are estimated using Monte Carlo (MC) simulation. The signal is extracted through a binned maximum likelihood fit of the sum of the expected background and signal to the \ptmiss distribution observed in data.

For this search we use data collected with the CMS detector in proton-proton collisions at a center-of-mass energy of 13\TeV in 2016--2018, which correspond to an integrated luminosity of $138\fbinv$. Tabulated results can be found in the HEPData record for this analysis~\cite{hepdata}.

\section{The CMS detector}

The CMS apparatus~\cite{CMS:2008xjf,CMS:2023gfb} is a multipurpose, nearly hermetic detector, designed to trigger on~\cite{CMS:2020cmk,CMS:2016ngn,CMS:2024aqx} and identify electrons, muons, photons, and (charged and neutral) hadrons~\cite{CMS:2020uim,CMS:2018rym,CMS:2014pgm}. Its central feature is a superconducting solenoid of 6\unit{m} internal diameter, providing a magnetic field of 3.8\unit{T}. Within the solenoid volume are a silicon pixel and strip tracker, a lead tungstate crystal electromagnetic calorimeter (ECAL), and a brass and scintillator hadron calorimeter (HCAL), each composed of a barrel and two endcap sections. Forward calorimeters extend the pseudorapidity coverage provided by the barrel and endcap detectors. Muons are reconstructed using gas-ionization detectors interleaved with the layers of the steel flux-return yoke outside the solenoid. More detailed descriptions of the CMS detector, together with a definition of the coordinate system used and the relevant kinematic variables, can be found in Refs.~\cite{CMS:2008xjf,CMS:2023gfb}. 

Events of interest are selected using a two-tiered trigger system. The first level, composed of custom hardware processors, uses information from the calorimeters and muon detectors to select events at a rate of around 100\unit{kHz} within a fixed latency of 4\mus~\cite{CMS:2020cmk}. The second level, known as the high-level trigger, consists of a farm of processors running a version of the full event reconstruction software optimized for fast processing, and reduces the event rate to a few kHz before data storage~\cite{CMS:2016ngn,CMS:2024aqx}.

\section{Event simulation}

Simulated samples of signal and background events are generated using various MC generators, and passed through a detailed simulation of the CMS detector implemented in \GEANTfour~\cite{Agostinelli:2002hh}. They are processed with the same version of the CMS event reconstruction software used for actual collision data samples. Due to the changes in the accelerator and detector configurations and conditions, separate samples are produced for each data-taking year.

Samples of signal events are simulated at leading order (LO) using \MGvATNLO 2.6.5~\cite{Alwall:2014hca}. The main parameters of the signal model are the coupling strength \coup of the $\chitwo\chione\yzero$ vertex, as well as the masses of the particles; all the other coupling strengths are set to 1. The complete simplified model contains additional couplings of the \yzero and \yone mediators to SM and dark-sector particles. In the benchmark interpretation considered here, these couplings and the corresponding interaction structure are fixed, while the four particle masses and the coupling \coup define the reduced parameter space explored in this analysis. In this search, we consider a range of values for that coupling strength and different combinations of particle masses, as listed in Table~\ref{sigmc}. The coupling strength affects the lifetime of the \chitwo particle and therefore impacts the displacement of its decay vertex with respect to the main interaction vertex. The mass of the \yone boson affects the kinematic properties of the signal, in particular the opening angle between the two \chitwo particles from the \yone decay and, consequently, the angular separation among the final-state light-flavour quarks from the two subsequent \yzero decays. 

The benchmark scenarios considered in this analysis are not intended to provide an exhaustive exploration of the reduced five-dimensional benchmark parameter space, spanned by the four particle masses and the coupling \coup, with the remaining couplings fixed as described above. Rather, they are chosen to probe the main phenomenological features of the signal. In particular, the selected benchmark points are chosen to probe different displacement regimes through variations of the coupling strength, as well as different boosted topologies through variations of the mediator mass. The \chitwo and \chione particle masses are kept fixed in these scans in order to isolate the impact of the displacement from that of the Lorentz boost on the reconstruction and tagging performance. This analysis represents a first study of this experimental signature in the boosted displaced topology, and the benchmark choices are intended as representative signal scenarios.

\begin{table}[!htbp]
  \centering
  \topcaption{Combinations of particle masses and the \coup probed in this search. The configurations in the upper block vary the coupling \coup for fixed masses of the new particles, whereas the configurations in the lower block vary the \yone boson mass for otherwise constant parameters. For all benchmark scenarios, the \yzero scalar mediator mass is fixed to $m_{\yzero}=1\GeV$. The quoted displacement $L_{xy}$ corresponds to a characteristic transverse decay length. The theoretical cross sections for the benchmark scenarios are listed in the last column.
  }
\begin{tabular}{llllll}
  \yone mass [\GeVns{}] & \chitwo mass [\GeVns{}] & \chione mass [\GeVns{}] & Coupling \coup & $L_{xy}$ [mm] & $\sigma_{\text{theory}}$ [pb] \\
\hline
300 & 100 & 30 & $1\ten{-7}$ & $\mathcal{O}(10)$ & 0.27\\
300 & 100 & 30 & $1\ten{-8}$ & $\mathcal{O}(10^{3})$ & 0.27\\
300 & 100 & 30 & $5\ten{-9}$ & $\mathcal{O}(10^{3})$ & 0.27\\
[\cmsTabSkip]
100 & 50 & 1 & $1\ten{-6}$ & $\mathcal{O}(0.1)$ & 0.17\\
300 & 50 & 1 & $1\ten{-6}$ & $\mathcal{O}(0.1)$ & 0.28\\
500 & 50 & 1 & $1\ten{-6}$ & $\mathcal{O}(0.1)$ & 0.15\\
\end{tabular}
\label{sigmc}
\end{table}

The dominant QCD multijet, \wj, and \zj background processes are generated at LO using \MGvATNLO. Top quark pair production and single top quark processes are simulated at next-to-LO accuracy in perturbative QCD using the \POWHEG 2.0~\cite{Nason:2004rx, Frixione:2007nw, Frixione:2007vw, Alioli:2009je, Alioli:2010xd, Campbell:2014kua} and \MGvATNLO generators, respectively. The decay of the top quark in the single top quark $t$-channel process is simulated with \MADSPIN~\cite{Artoisenet:2012st}.

The parton shower and hadronization are modeled with \PYTHIA~8.230 for all simulated samples, using the CP5 tune of \PYTHIA parameters~\cite{CMS:2019csb} to describe the underlying event. All samples were generated using the NNPDF3.1 next-to-next-to-LO parton distribution functions (PDFs)~\cite{Ball:2017nwa}. We overlay the main simulated interaction with additional simulated inelastic proton-proton interactions in the same or adjacent bunch crossings (pileup), and reweight events to match the measured number of pileup interactions in the data.

\section{Event reconstruction and selection}
\label{obj}

 The global event reconstruction, also called particle-flow (PF) event reconstruction~\cite{CMS:2017yfk}, aims to reconstruct and identify each individual particle in an event, with an optimized combination of all subdetector information. In this process, the identification of the particle type (photon, electron, muon, charged hadron, neutral hadron) plays an important role in the determination of the particle direction and energy. Photons are identified as ECAL energy clusters not linked to the extrapolation of any charged-particle trajectory to the ECAL. Electrons are identified as a primary charged-particle track and potentially many ECAL energy clusters corresponding to this track's extrapolation to the ECAL and to possible bremsstrahlung photons emitted along the way through the tracker material. Muons are identified as tracks in the central tracker consistent with either a track or several hits in the muon system, and associated with calorimeter deposits compatible with the muon hypothesis. Charged hadrons are identified as charged-particle tracks not linked to electrons or muons. Finally, neutral hadrons are identified as HCAL energy clusters not linked to any charged-hadron trajectory, or as a combined ECAL and HCAL energy excess with respect to the expected charged-hadron energy deposit.

Photon energies are measured using the ECAL. For electrons, the momentum measured by the tracker at the primary interaction vertex is combined with the corresponding ECAL cluster and the summed contribution from associated bremsstrahlung photons. Muon momenta are obtained from the reconstructed tracks. For charged hadrons, the track momentum is combined with the associated ECAL and HCAL measurements, including corrections for the calorimeter response to hadronic showers. Neutral-hadron energies are determined from the corrected ECAL and HCAL measurements.

Hadronic jets are clustered from the PF candidates using the infrared- and collinear-safe anti-\kt algorithm~\cite{Cacciari:2008gp,Cacciari:2011ma} with a distance parameter of 0.4 or 0.8 (AK4 or AK8 jets, respectively). The pileup-per-particle identification (PUPPI) algorithm~\cite{Bertolini:2014bba,Sirunyan:2020foa} is used to mitigate pileup effects at the reconstructed-particle level. Its effect on the reconstruction of displaced signatures is included in the determination of the jet tagging efficiency. A potentially degraded performance is partially recovered by supplementing the PUPPI jet constituents with explicitly reconstructed secondary-vertex information in the GNN input. Jet energy corrections are derived from simulation to bring the measured response of jets to that of particle-level jets on average. In situ measurements of the momentum balance in dijet, photon+jet, \PZ{}+jet, and multijet events are used to account for any residual differences in the jet energy scale (JES) between data and simulation~\cite{CMS:2016lmd}. Additional selection criteria are applied to each jet to remove jets potentially dominated by anomalous contributions from various subdetector components or reconstruction failures. Small-radius AK4 jets are only used in the online trigger reconstruction for the calculation of the scalar transverse momentum sum of jets.

The missing transverse momentum vector \ptvecmiss is computed as the negative vector sum of the transverse momenta of all the PF candidates in an event, and its magnitude is denoted as \ptmiss~\cite{CMS:2019ctu}. The \ptvecmiss is modified to account for corrections to the energy scale of the reconstructed jets in the event. The PUPPI algorithm is also applied to the \ptvecmiss observable to reduce the pileup dependence. Anomalous high-\ptmiss events can be due to a variety of reconstruction failures, detector malfunctions, or noncollision backgrounds. Event filters~\cite{CMS:2019ctu} are applied to reject such events with spurious \ptmiss.

A combination of triggers is employed requiring events to have high values of the scalar \pt sum of all AK4 jets in an event or AK8 jets with high \pt or trimmed mass~\cite{Krohn:2009th}. The AK8 jets reconstructed offline are required to have $\pt > 200\GeV$ and $\abs{\eta} < 2.5$, and pass the tight jet identification requirement~\cite{CMS-PAS-JME-16-003}. In this search, we require the presence of at least two AK8 jets satisfying the aforementioned criteria to select events compatible with the expected signature of one ISR jet and the decay of a Lorentz-boosted \yone boson into a \chitwochitwo pair. The \pt-leading AK8 jet in the event is additionally required to have $\pt > 500\GeV$ to ensure high trigger efficiency.

A GNN-based classifier is used to identify the signal AK8 jets based on their substructure. The interaction network architecture~\cite{gnn} is utilized: a graph is formed to represent the process, where the objects of interest---namely the jet constituents (particles) characterized by their four-momenta---are represented by nodes and are connected to each other via edges. The interactions and correlations between the nodes are taken into account by the classifier (tagger) trained to identify the signal jet.

The graph is constructed using the reconstructed PF candidates associated with the AK8 jet, supplemented with secondary vertex information to enhance sensitivity to displaced topologies. Each node is characterized by kinematic and tracking-related observables, while the graph edges encode pairwise relations between constituents. The complete set of input features is summarized in Table~\ref{input_vars}. A fully connected graph topology is employed, allowing the interaction network to learn nonlocal correlations among jet constituents. The relational information is embedded through a sequence of fully connected layers acting on the node and edge representations, followed by aggregation operations used to construct a jet-level representation and classification score.

The tagger is trained using simulated signal jets and QCD multijet background jets, which constitute the dominant background process in the analysis. The signal samples used for training are presented in Table~\ref{sigmc} and span a wide range of mediator masses and lifetimes, ensuring sensitivity across different displacement regimes and kinematic configurations. Background jets are selected to match the kinematic properties of the signal jets, in particular in transverse momentum and pseudorapidity, to avoid biases in the classifier response.

\begin{table}[!ht]
  \centering
  \topcaption{Input features used by the GNN-based jet tagger.}
  \small
  \begin{tabular}{
    >{\raggedright\arraybackslash}p{0.36\textwidth}
    >{\raggedright\arraybackslash}p{0.58\textwidth}
  }
    \noalign{\vskip 3pt}
    Input features & Description \\
    \hline
    \noalign{\vskip 3pt}

    $p_{\mathrm{T}}^{\mathrm{cand}}/p_{\mathrm{T}}^{\mathrm{jet}}$,
    $E^{\mathrm{cand}}/E^{\mathrm{jet}}$
    &
    Transverse momentum and energy of the charged PF candidate relative to the AK8 jet
    \\\noalign{\vskip 3pt}

    $\Delta\eta(\mathrm{cand},\mathrm{jet})$,
    $\Delta\phi(\mathrm{cand},\mathrm{jet})$,
    $\Delta R(\mathrm{cand},\mathrm{jet})$
    &
    Position of the charged PF candidate relative to the AK8 jet axis
    \\\noalign{\vskip 3pt}

    $\Delta R_{\min}(\mathrm{cand},\mathrm{SV})$,
    $\Delta R_i(\mathrm{cand},\mathrm{subjet}_i)$
    &
    Angular distance to the closest secondary vertex and to the two leading subjets
    \\\noalign{\vskip 3pt}

    $d_z$, $d_z/\sigma(d_z)$,
    $d_{xy}$, $d_{xy}/\sigma(d_{xy})$
    &
    Longitudinal and transverse impact parameters and their significances
    \\\noalign{\vskip 3pt}

    $\chi^2_{\mathrm{track}}/\mathrm{ndf}$,
    track quality
    &
    Track-fit quality information
    \\\noalign{\vskip 3pt}

    Track covariance matrix elements
    &
    Covariances involving $p_{\mathrm{T}}$, $\eta$, $\phi$, $d_{xy}$,
    $d_z$, and the track dip angle, defined as the angle between the
    track momentum and the transverse plane
    \\\noalign{\vskip 3pt}

    $\eta_{\mathrm{rel}}$,
    $p_{\mathrm{T,rel}}/p$,
    $p_{\parallel}/p$
    &
    Track kinematic properties relative to the AK8 jet axis
    \\\noalign{\vskip 3pt}

    $\mathrm{SIP}_{\mathrm{2D}}$,
    $\mathrm{SIP}_{\mathrm{2D}}/\sigma(\mathrm{SIP}_{\mathrm{2D}})$,
    $\mathrm{SIP}_{\mathrm{3D}}$,
    $\mathrm{SIP}_{\mathrm{3D}}/\sigma(\mathrm{SIP}_{\mathrm{3D}})$
    &
    Signed two- and three-dimensional impact parameters and their significances
    \\\noalign{\vskip 3pt}

    $d(\mathrm{track},\mathrm{jet\ axis})$
    &
    Minimum distance between the track trajectory and the jet axis
    \\

    \hline
    \noalign{\vskip 3pt}
    $p_{\mathrm{T}}^{\mathrm{SV}}/p_{\mathrm{T}}^{\text{jet}}$,
    $E^{\mathrm{SV}}/E^{\text{jet}}$
    &
    Transverse momentum and energy of the secondary vertex relative to the AK8 jet
    \\\noalign{\vskip 3pt}

    $\Delta\eta(\mathrm{SV},\text{jet})$,
    $\Delta\phi(\mathrm{SV},\text{jet})$,
    $\Delta R(\mathrm{SV},\text{jet})$
    &
    Position of the secondary vertex relative to the AK8 jet axis
    \\\noalign{\vskip 3pt}

    $p_{\mathrm{T}}^{\mathrm{SV}}$,
    $m_{\mathrm{SV}}$,
    $N_{\mathrm{tracks}}^{\mathrm{SV}}$,
    $\chi^2_{\mathrm{SV}}/\mathrm{ndf}$
    &
    Secondary-vertex kinematic properties, mass, track multiplicity, and fit quality
    \\\noalign{\vskip 3pt}

    $\cos\theta_{\mathrm{SV,PV}}$
    &
    Cosine of the angle between the line connecting the primary and secondary
    vertices and the momentum direction of the particles associated with the
    secondary vertex
    \\\noalign{\vskip 3pt}

    $d_{xy}^{\mathrm{SV}}$,
    $d_{xy}^{\mathrm{SV}}/\sigma(d_{xy}^{\mathrm{SV}})$,
    $d_{\mathrm{3D}}^{\mathrm{SV}}$,
    $d_{\mathrm{3D}}^{\mathrm{SV}}/\sigma(d_{\mathrm{3D}}^{\mathrm{SV}})$
    &
    Transverse and three-dimensional displacements of the secondary vertex and their significances
    \\

    \hline
    \noalign{\vskip 3pt}
    $\Delta\phi(j_1,j_2)$,
    $\Delta R(j_1,j_2)$
    &
    Angular separation between the two leading AK8 jets
    \\

  \end{tabular}
  \label{input_vars}
\end{table}

Up to 60 charged PF candidates and 5 secondary vertices are retained for each AK8 jet. The charged PF candidates are ordered by decreasing transverse impact parameter significance, while the secondary vertices are ordered by decreasing transverse displacement significance. Collections containing fewer objects are zero-padded to the corresponding maximum size.

The background jets used for training are selected to match the \pt and $\eta$ distributions of the signal jets in order to avoid biases in the classifier response arising from differences in the basic kinematic properties of the jets. No additional requirement on the jet mass or jet substructure is imposed before training in order to avoid sculpting the background sample or introducing trivial differences between the training classes.

The hyperparameters of the network architecture and training procedure were optimized using the expected signal-to-background discrimination performance on statistically independent validation samples. The training performance was validated by comparing the classifier response in independent simulated samples and in background-enriched CR in data.

Overall, the GNN tagger exploits both kinematic and topological information of jet constituents to enhance the discrimination between signal-like jets, characterized by multiprong and potentially displaced substructure, and background jets arising from SM processes.

\begin{figure}[ht!]
  \centering
    \includegraphics[width=0.49\textwidth]{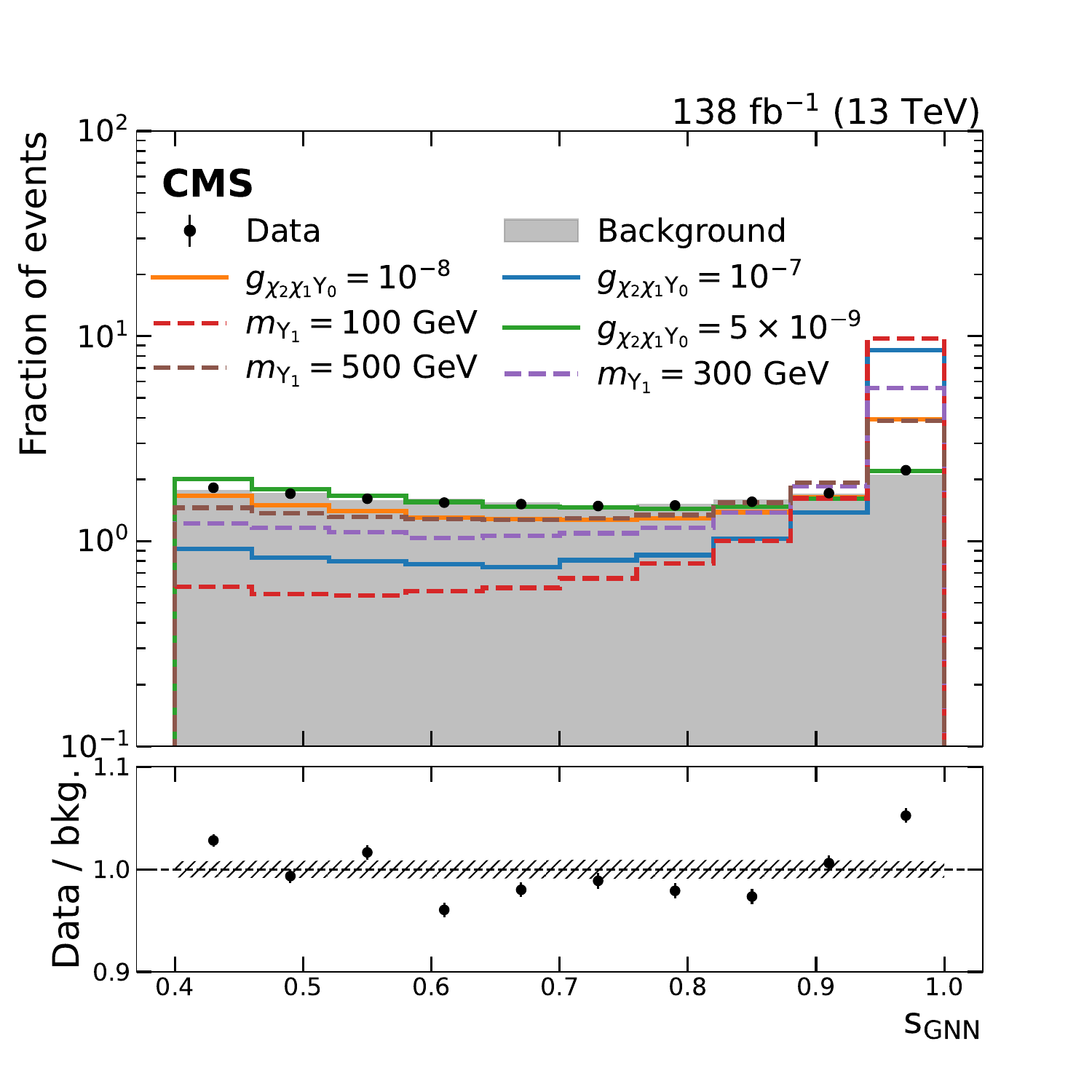}
    \caption{Distribution of \sgnn in the SR before imposing any requirement on \sgnn for the range of values considered in this analysis. The data are shown as black markers with vertical bars indicating the statistical uncertainty. Signal processes are shown as solid lines, while the total background, corresponding to the sum of all considered background contributions estimated from simulation, is represented by the filled histogram. The lower panel shows the ratio of the data and the total background prediction, and the hatched area represents the total uncertainty in the background. The different signal points shown in the figure are listed in Table~\ref{sigmc}. All distributions are normalized to unity.}
\label{fig:gnn_score}
\end{figure}

The distribution of the tagger output score \sgnn in the SR before imposing any requirement on \sgnn is shown in Fig.~\ref{fig:gnn_score} for the range of values considered in this analysis. All histograms in the figure are normalized to the same area. The observed discrepancy between data and simulation arises from imperfect QCD background modeling and does not affect the analysis since the QCD background contribution in the analysis is derived from a dedicated CR in data.
Signal distributions are represented by solid lines, while the total simulated background (corresponding to the sum of all background processes in the analysis) is shown as a filled histogram.
It can be observed that the signal distributions are well separated from the background. However, the GNN performance degrades for signals with large displacement, as highly displaced objects are more difficult to reconstruct. In addition, the performance decreases for higher-mass signals since their decay products are less likely to be reconstructed in a single AK8 jet. The area under the receiver operating characteristic curve of the tagger ranges from approximately 0.88 for the most displaced and heaviest signal to about 0.95 for the lightest signal.

Of the two leading AK8 jets in each event, the one with the highest tagger score is selected as the candidate AK8 jet. This jet is further required to have an ungroomed PUPPI AK8 jet mass exceeding 40\GeV and a tagger score $\sgnn > 0.8$. 

A minimum \ptmiss of 200\GeV is required, which suppresses the SM backgrounds without genuine \ptmiss, such as the dominant background from QCD multijet events. Events that contain a reconstructed lepton passing the following criteria are rejected. Electrons are required to have $\pt > 10\GeV$, $\abs{\eta} < 2.5$, and pass the set of loose identification criteria outlined in Ref.~\cite{CMS:2020uim}. Muons are required to have $\pt > 10\GeV$, $\abs{\eta} < 2.1$, pass the set of loose identification criteria outlined in Ref.~\cite{CMS:2018rym}, and have a PF-based combined relative isolation value~\cite{CMS:2017yfk} of less than 0.4. Hadronically decaying tau leptons are identified with the \textsc{DeepTau} algorithm~\cite{Tumasyan_2022}, and are required to have $\pt > 20\GeV$ and $\abs{\eta} < 2.3$. Events passing all aforementioned requirements enter the SR of this analysis.

\section{Background estimation}
\label{Background Estimation}
\label{bkg}

Events arising from SM QCD multijet, \wj, and \zj production constitute the primary background process in this analysis. The \zj events may pass the event selection via the $\PZ\to\nu\nu$ decay, which produces significant \ptmiss. The QCD multijet background dominates at low \ptmiss, while the \zj and \wj processes become increasingly relevant in the high-\ptmiss tail, which is also the region most sensitive to the signal. Three dedicated CRs are defined to estimate the contribution of each of these main backgrounds, whereas subdominant processes are estimated directly from simulation. The CRs are obtained from the SR selection by modifying the $\sgnn$ requirement and/or imposing lepton multiplicity requirements, as appropriate for the targeted background process; their detailed definitions are given below.

The QCD multijet background is estimated using a CR designed to closely mirror the SR, except for the requirement on the score of the GNN tagger. The QCD CR is defined by a GNN score window of $0.4 < \sgnn < 0.8$, ensuring orthogonality to the SR ($\sgnn > 0.8$), while maintaining sufficient statistical power and minimizing signal contamination. 
The overall yield of QCD multijet events in the SR and all CRs is determined in situ from a simultaneous fit to the data \ptmiss or hadronic recoil distributions in these four regions. In each \ptmiss or hadronic recoil bin, the QCD CR is used to estimate the yields of the QCD multijet distributions in the non-QCD CRs and in the SR. Since all other selection requirements are shared between the QCD CR and the SR, modeling effects common to the two regions largely cancel in the transfer factor; no significant correlation between $\sgnn$ and \ptmiss is observed in simulation. The per-bin yield in the QCD CR is linked to the yield in the corresponding bin in each of the three other regions via a transfer factor (TF). Each TF is defined as the ratio of the simulated QCD multijet yield in the respective region and in the QCD CR in bin $i$. For the SR, and analogously for each of the two non-QCD CRs, the TF is given by
\begin{equation}
\mathrm{TF}_i^{\text{QCD, SR}} = \frac{N_i^{\text{QCD, SR}}}{N_i^{\text{QCD, CR}}}.
\end{equation}
The number of QCD multijet events in each bin of the SR is then defined as
\begin{equation}
N_{i\text{,\,data}}^{\text{QCD, SR}} = \mathrm{TF}_i^{\text{QCD, SR}}\,N_{i\text{,\,data}}^{\text{QCD, CR}},
\end{equation}
where $N_{i\text{,\,data}}^{\text{QCD, SR}}$ is left unconstrained in the final fit, effectively being determined by the QCD CR data via the TF. The TF in each bin is determined from simulated QCD multijet events and is subject to statistical fluctuations due to the limited number of simulated events, as well as experimental uncertainties. 

The \wj background, predominantly entering the SR if the charged lepton from the $\PW\to\ell\nu$ decay does not pass the criteria employed to reject events with leptons, is estimated using events containing exactly one lepton passing those criteria and an AK8 jet with a GNN score in the range $0.4 < \sgnn < 0.8$. The choice of the $\sgnn$ interval provides a compromise between maintaining sufficient statistical precision, particularly at high \ptmiss, and keeping the jet selection close to that in the SR. This CR is fitted simultaneously with the SR and the other CRs for the signal extraction. A separate unconstrained yield parameter and TFs, defined analogously to those used for the QCD multijet background estimation, are employed to extrapolate from the \wj CR to the SR and the other CRs. 

The \zvvj process, dominant at high \ptmiss, is estimated using $\PZ(\to\Pell\Pell)+\text{jets}$ events as a proxy, where \Pell denotes an electron or muon. The hadronic recoil $\Vec{U}$, defined as
\begin{equation}
\Vec{U} = \ptvecmiss + \sum_{\Pell} \ptvec,
\end{equation}
is used to emulate the \ptmiss distribution of \zvvj events. Events are required to contain two same-flavor opposite-sign leptons satisfying the standard identification and isolation criteria described in Section~\ref{obj}, with an invariant mass in the range $60 < m_{\Pell\Pell} < 120\GeV$, and at least one AK8 jet with $\pt > 500\GeV$. Because of the smaller size of the $\PZ(\to\Pell\Pell)+\text{jets}$ data sample, no additional requirement on $\sgnn$ is imposed, in order to preserve sufficient statistical precision in the hadronic recoil bins. The distribution of \zvvj events in the SR and the other CRs is predicted from the distribution of the magnitude of the hadronic recoil in this CR via TFs and an unconstrained yield parameter similarly to the estimation of the QCD multijet and \wj backgrounds.

\section{Systematic uncertainties}
\label{sec:syst}

Systematic uncertainties arise from imperfect knowledge of detector effects, theoretical modeling, and corrections derived from data. These can affect both the normalization and the shape of the discriminating observables used in the maximum likelihood fits described in Section~\ref{results}.
Each source of uncertainty is modeled as a nuisance parameter: uncertainties affecting only the rate of a specific process use log-normal constraints, while shape-dependent ones are implemented via a polynomial interpolation with Gaussian prior distributions. Correlations between years of data taking and processes are imposed, as appropriate.

The integrated luminosities for the 2016, 2017, and 2018 data-taking years have 1.2--2.5\% individual uncertainties~\cite{CMS-LUM-17-003,CMS-PAS-LUM-17-004,CMS-PAS-LUM-18-002}, while the overall uncertainty for the 2016--2018 period is 1.6\%. 
An uncertainty of 0.5\% is assigned to the electron, muon, and tau lepton veto efficiencies. The uncertainties associated with the trigger efficiency scale factors are negligible.

Variations of the JES and resolution (JER) by $\pm1$ standard deviation are propagated to all relevant physics objects.
The resulting variations are treated as affecting only the yields of the fitted distributions.
An uncertainty related to the unclustered energy contribution to the \ptmiss is considered. The unclustered energy corresponds to PF candidates not associated with reconstructed jets. Its effect is evaluated by varying the transverse momenta of these candidates within their uncertainties and propagating the variation to the \ptmiss observable. The resulting change in the \ptmiss distribution is treated as a shape uncertainty, fully correlated across all bins and processes.

The efficiency of the GNN-based jet tagger is determined using a reweighting method based on the Lund jet plane (LJP)~\cite{ljp_paper}.
This approach corrects the substructure modeling of multiprong, displaced large-radius jets by applying per-prong corrections based on the ratio of LJP densities in data and simulation. The LJP reweighting method has previously been used mainly for prompt multiprong jet signatures; in this analysis it is extended for the first time to displaced large-radius jets. The displacement of the (\chitwo) decay can affect the reconstructed jet substructure through changes in the PF-candidate composition, track association, secondary-vertex reconstruction, and subjet-to-parton matching. The applicability of the correction is therefore validated in simulated signal samples with different lifetimes, and uncertainties are assigned to cover the residual limitations of applying the method to displaced topologies.
The resulting per-event scale factors are applied to signal samples and do not affect the shape of the \ptmiss distribution.
The scale factors derived for representative signal parameter configurations, together with the absolute reconstruction efficiency, are summarized in Table~\ref{tab:tagger_sf}.

\begin{table}[!ht]
  \centering
  \topcaption{Tagger efficiency scale factors and absolute reconstruction efficiency derived for the representative signal parameters.}
  \begin{tabular}{l*{6}{c}}
    & \multicolumn{3}{c}{\coup} & \multicolumn{3}{c}{\yone mass [\GeVns{}]} \\
    & $1\times10^{-7}$ & $1\times10^{-8}$ & $5\times10^{-9}$ & 100 & 300 & 500 \\
    \hline
    Scale factor & 0.97 & 0.99 & 1.01 & 1.01 & 0.93 & 0.89 \\
    Uncertainty & 60\% & 50\% & 50\% & 50\% & 60\% & 70\% \\
    Absolute reconstruction efficiency & 92\% & 91\% & 89\% & 94\% & 92\% & 93\% \\
  \end{tabular}
  \label{tab:tagger_sf}
\end{table}

The total uncertainty in the tagger efficiency is dominated by subjet matching effects in the LJP correction procedure. These are grouped into three categories: jets with generator-level quarks near the geometrical AK8 jet boundary, unmatched subjets, or ambiguous subjet-quark associations. Following the methodology of Ref.~\cite{ljp_paper}, the matching-related uncertainties are evaluated from changes in the tagger efficiency scale factors obtained by repeating the LJP correction with different numbers of subjets and varying the weights of events containing generator-level quarks that remain unmatched to subjets even after these changes, while preserving the overall sample normalization. The dominant contribution to the 50--70\% uncertainty in the tagger efficiency scale factors arises from the strong collimation of the two quarks from each light \yzero boson decay, which hinders reliable subjet-quark matching and the application of per-prong LJP corrections, and thus reflects a limitation associated with the signal topology rather than an effect unique to displaced decays.

Uncertainties are assigned to each bin of the TF for the QCD, \wj, and \zvvj CRs. To account for the statistical precision of the simulated samples, each bin is assigned an independent uncertainty, allowing the TF for that bin to fluctuate within its statistical uncertainty. In addition to these per-bin statistical components, experimental uncertainties related to the JES and JER are propagated to the TFs. These variations are treated as fully correlated across all bins, as they arise from global detector effects.

Table~\ref{tab:syst_summary} summarizes the systematic uncertainties considered in this analysis. After the GNN tagger uncertainty, the largest contribution is the uncertainty in the bin-by-bin TF.
\begin{table}[ht]
  \centering
  \topcaption{Summary of the systematic uncertainties considered in this analysis. The second column details whether the uncertainty affects the shape or the normalization (norm.) of the fitted distribution. Uncertainty in the third column reports the effect of a given uncertainty on the normalization of the affected processes.}
\begin{tabular}{llll}
Source & Type & Effect [\%] & Comment \\
\hline
Integrated luminosity & norm. & 1--3 & \\
Lepton veto & norm. & 0.5 & \\
AK8 JES & norm. & 2 & \\
AK8 JER & norm. & 2 & \\
GNN tagger efficiency & norm. & 50--70 & Affects signal only \\
QCD TF & norm. & 5--8 & Per-bin statistical uncertainty \\
\wj TF & norm. & 5--8 & Per-bin statistical uncertainty \\
$\PZ \to \Pgn\Pgn$ TF & norm. & 5--8 & Per-bin statistical uncertainty \\
\ptmiss unclustered & shape & 1--3 & Fully correlated across all bins \\
\ptmiss JES/JER & shape & 1--3 & Fully correlated across all bins \\
\end{tabular}
\label{tab:syst_summary}
\end{table}

\section{Results}
\label{results}

A binned maximum likelihood fit to the \ptmiss distribution in the SR is performed to test for the presence of a signal. The results are interpreted using the modified frequentist \CLs~\cite{Junk:1999kv,Read:2002hq} criterion with the profile likelihood ratio as the test statistic, and upper limits at 95\% confidence level (\CL) are set on the signal strength $\mu=\sigma/\sigma_{\text{theory}}$ using the asymptotic approximation~\cite{Cowan_2011}, as implemented in the CMS statistical analysis tool \textsc{Combine}~\cite{CMS:2024onh}. 

Separate distributions for each year of data taking are fitted simultaneously to obtain the final results, taking into account possible correlations between the systematic uncertainties, as described in Section~\ref{sec:syst}.
The QCD multijet, \wj, and \zj contributions are estimated from dedicated CRs described in detail in Section~\ref{bkg}. All other backgrounds are estimated directly from simulation. No significant excess over the SM expectation is observed.

\begin{figure}[!ht]
  \centering
    \includegraphics[width=0.475\textwidth]{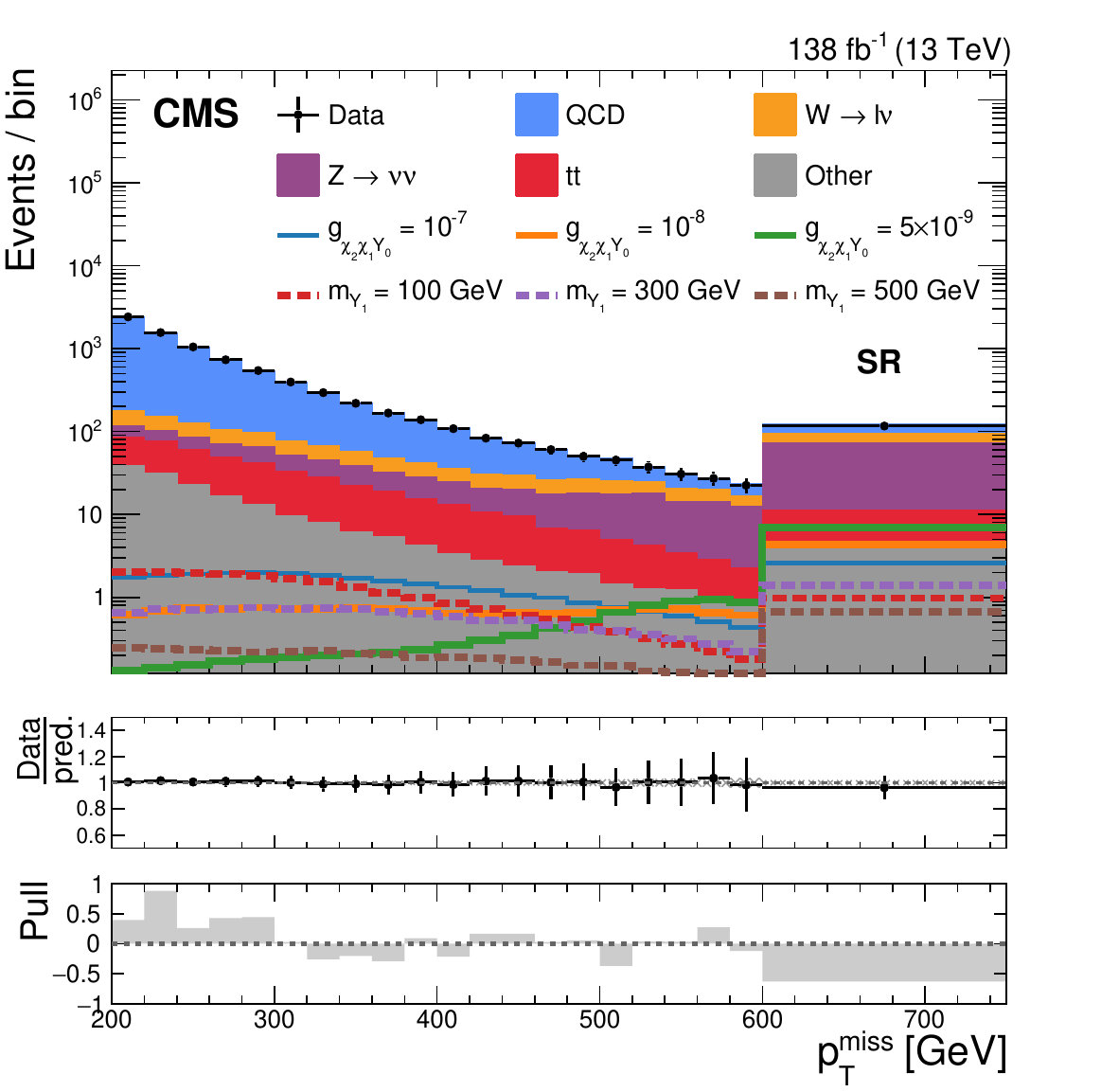}
    \includegraphics[width=0.475\textwidth]{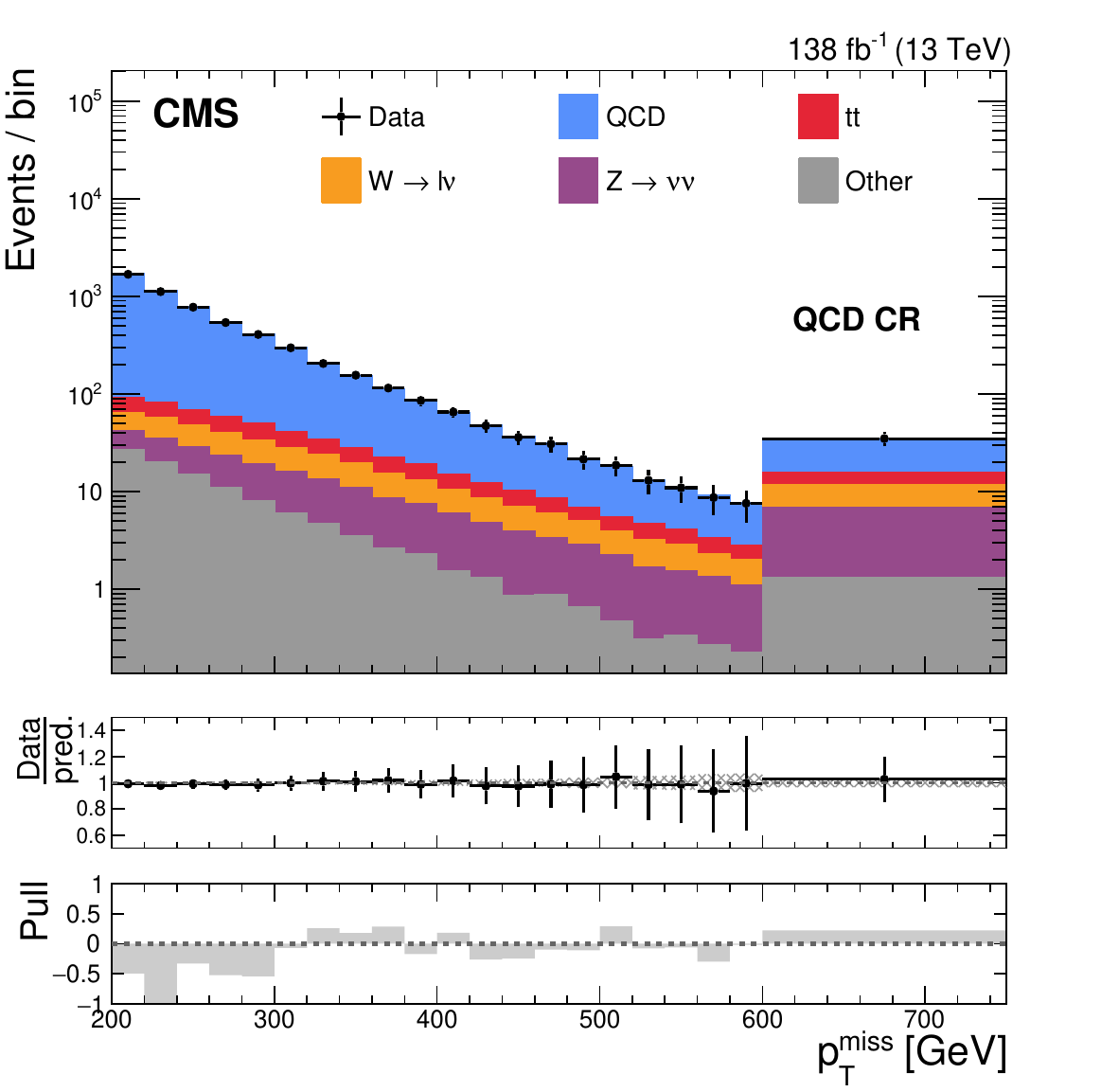}
    \includegraphics[width=0.475\textwidth]{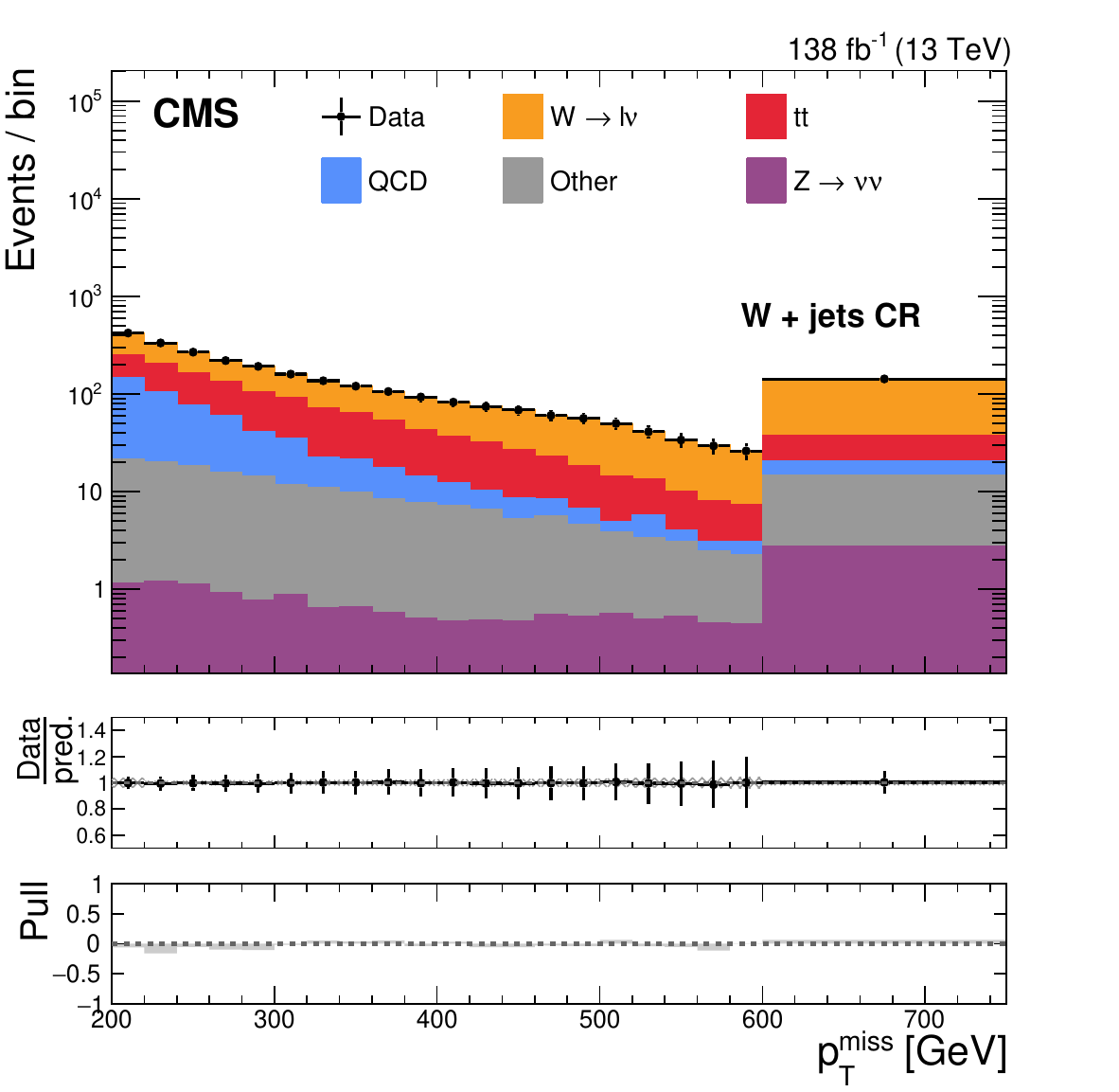}
    \includegraphics[width=0.475\textwidth]{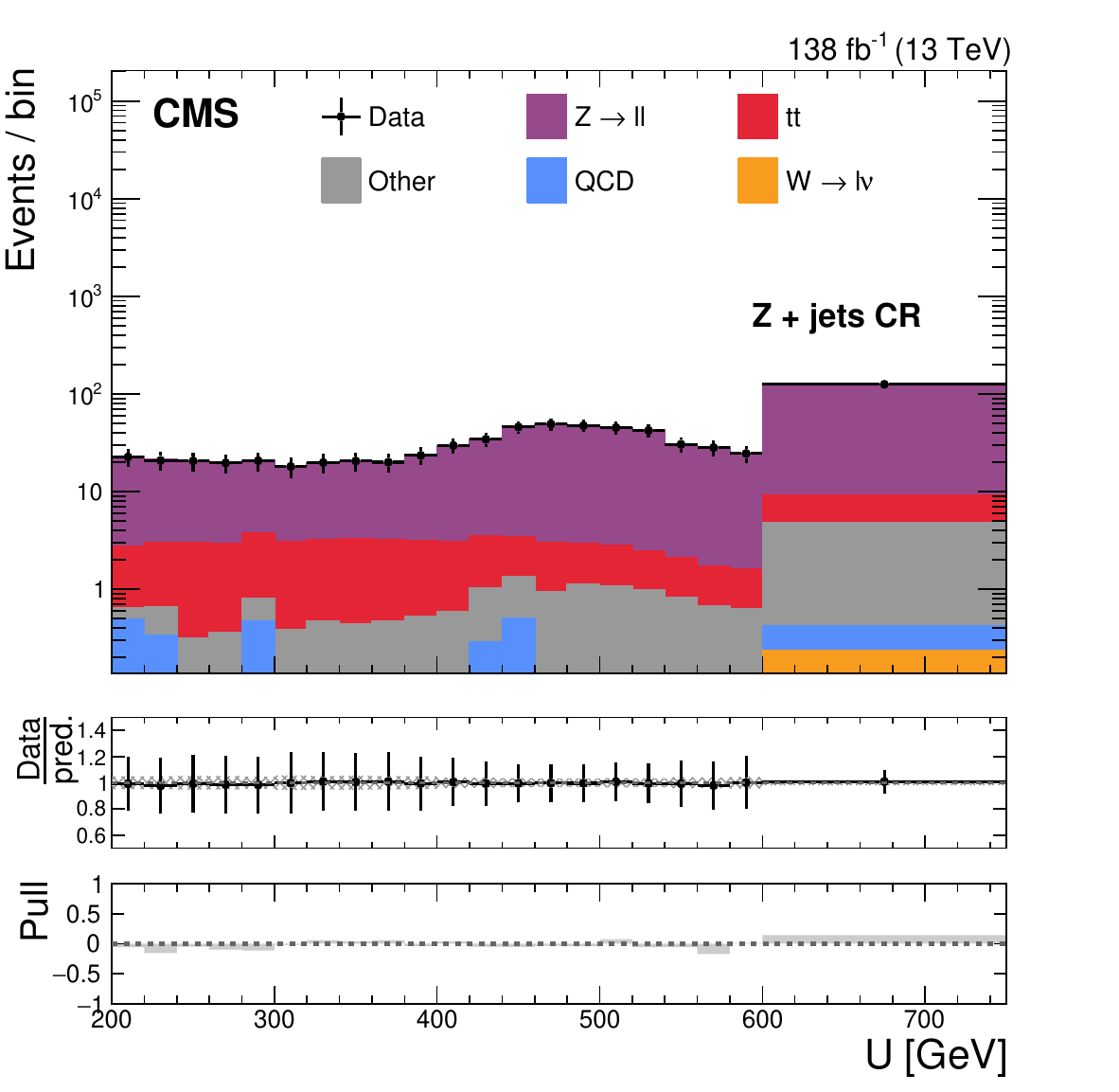}
    \caption{Postfit \ptmiss distributions in the SR (upper left), QCD CR (upper right), \wj CR (lower left), and postfit distribution of the magnitude of the hadronic recoil in the \zj CR (lower right). The data are shown as black markers with vertical bars indicating the statistical uncertainty. The SM expectation is shown as stacked histograms. In the SR, the prefit expected signal contributions are displayed as dashed lines for different signal parameter settings (different colors); the complete model parameter configurations corresponding to the signal hypotheses shown in the SR are listed in Table~\ref{sigmc}. In the middle panels, the ratio of the data to the postfit SM prediction is shown, with the total uncertainty in the latter represented as a hatched area. In the lower panels, the difference between data and SM expectation divided by the uncertainty in that difference (pull) is reported. The last bin includes the overflow events.}
    \label{fig:postfit}
\end{figure}

Figure \ref{fig:postfit} shows the distributions of \ptmiss in the SR and CRs after a background-only fit to the data and combining all data-taking periods for illustration. The per-bin normalizations of the dominant QCD multijet, $\wj$, and $\zj$ backgrounds are constrained primarily by the corresponding CR data. The transfer factors used to extrapolate between regions are derived from simulation and are profiled within their per-bin statistical and correlated experimental uncertainties. The prefit signal distributions for the different mass and coupling hypotheses are also shown in the SR to illustrate the expected signal shapes, given that no significant excess is observed. 

\begin{figure}[!ht]
  \centering
    \includegraphics[width=0.49\textwidth]{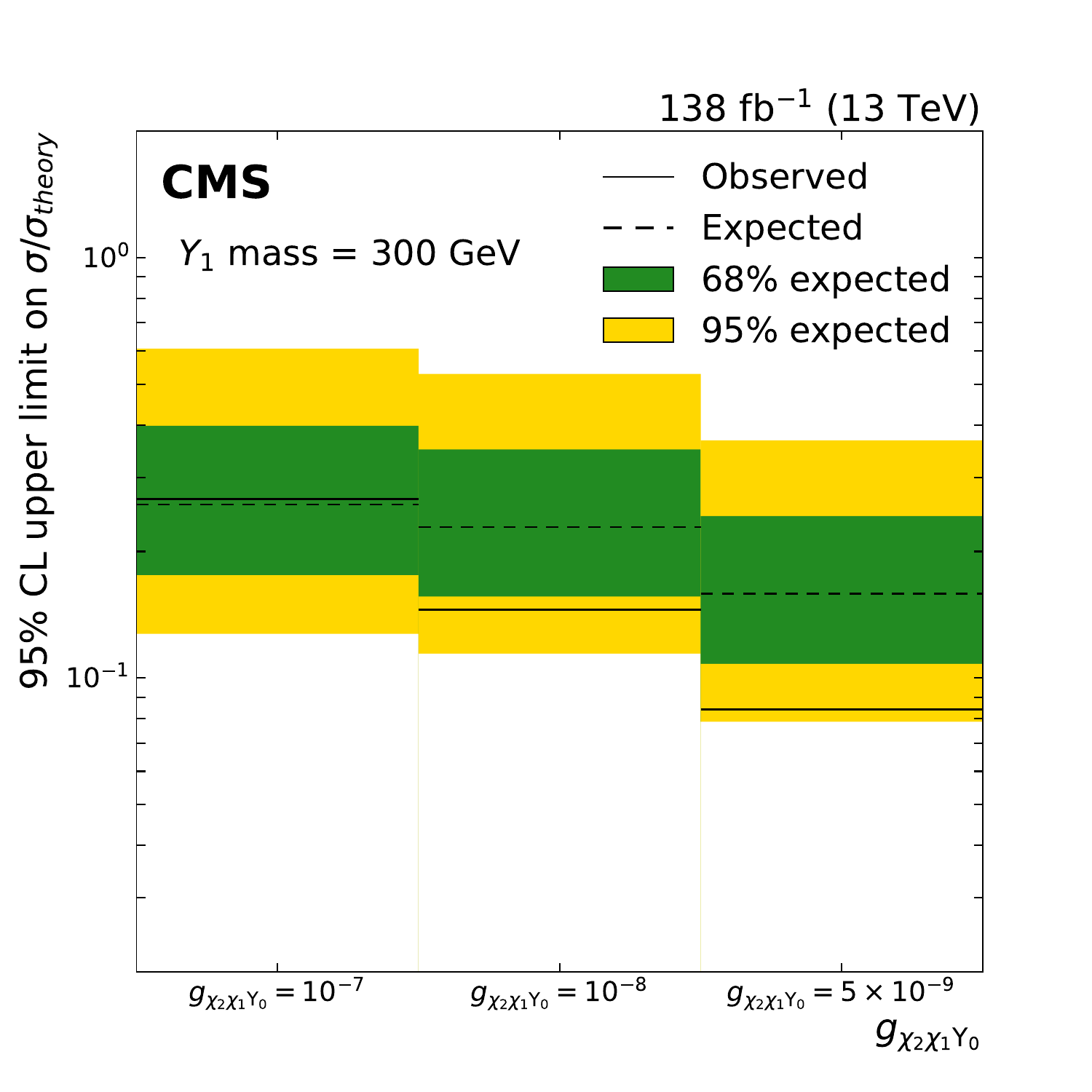}
    \includegraphics[width=0.49\textwidth]{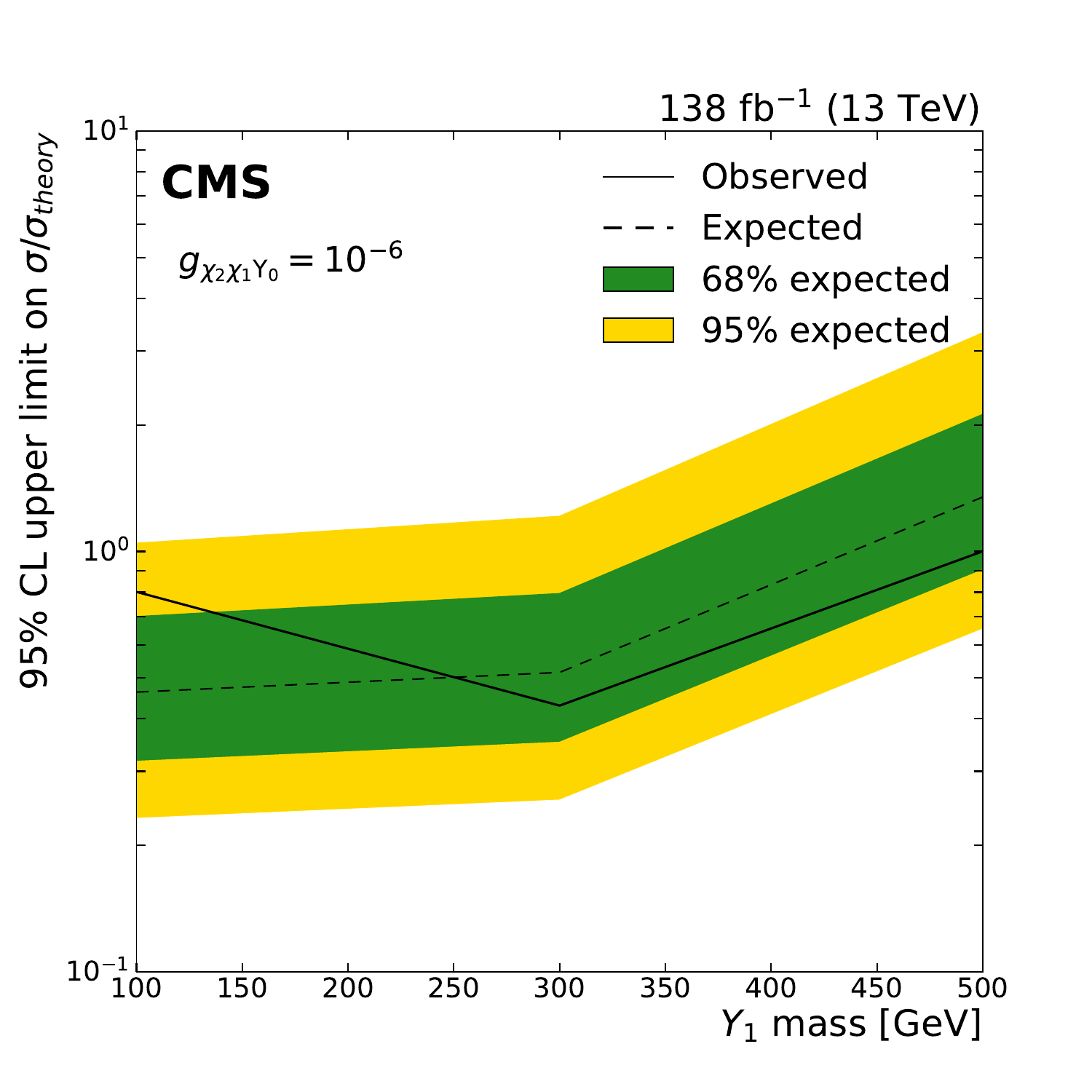}
    \caption{Upper limits at 95\% \CL on the signal strength for different scenarios of the coupling \coup (left) and varying \yone masses (right). The observed (expected) limits are shown as a solid (dashed) black line and the inner (green) band and the outer (yellow) band indicate the regions containing 68\% and 95\%, respectively, of the distribution of limits expected under the background-only hypothesis. The $\sigma_{\text{theory}}$ values for the signal are reported in Table~\ref{sigmc}.
    } 
\label{fig:upper_lim}
\end{figure}

Upper limits at 95\% \CL on \coup and the mass of the \yone mediator particle are derived for the signal parameter configurations given in Table~\ref{sigmc}, and are shown in Fig.~\ref{fig:upper_lim} (left and right, respectively). The observed and expected upper limits agree within two standard deviations. The good postfit agreement observed in the CRs is expected because the dominant QCD, \wj, and \zj backgrounds are constrained directly from data in the simultaneous fit through dedicated CRs, with their normalizations and transfer factors determined in situ. The observed limits in Fig.~\ref{fig:upper_lim} are obtained from the full profile likelihood fit, including the SR and CRs, all bins and data-taking years, correlations, and systematic uncertainties. Consequently, small residual differences between the data and the postfit background prediction in the SR, even when below one standard deviation per bin, can shift the observed limit with respect to the median expected limit if they occur coherently in bins with high signal sensitivity. These results constrain the parameter space of the model, excluding at 95\% \CL the signal hypotheses corresponding to the tested values of the mediator mass and coupling strength shown in Fig.~\ref{fig:upper_lim}.

\section{Summary}
The first search for a pair of nonprompt dark matter candidates in the Lorentz-boosted topology, characterized by a large-radius jet and large missing transverse momentum, has been presented. The analysis uses data collected in proton-proton collisions at $\sqrt{s}=13\TeV$, corresponding to an integrated luminosity of $138\fbinv$. Signal candidates are selected by requiring the presence of a large-radius jet that is identified as signal-like using a tagger based on a graph neural network that exploits jet substructure and secondary vertex information. The dominant standard model backgrounds from quantum chromodynamics multijet production, \zvvj, and \wj processes are estimated from data using dedicated control regions. The missing transverse momentum spectrum is analyzed to search for a potential signal above the expected background. No significant excess over the standard model expectation is observed. Upper limits at 95\% confidence level are set on the signal strength as functions of either the mediator mass or the coupling strength of the $\chitwo\chione\yzero$ vertex. 

\begin{acknowledgments}
We congratulate our colleagues in the CERN accelerator departments for the excellent performance of the LHC and thank the technical and administrative staffs at CERN and at other CMS institutes for their contributions to the success of the CMS effort. In addition, we gratefully acknowledge the computing centres and personnel of the Worldwide LHC Computing Grid and other centres for delivering so effectively the computing infrastructure essential to our analyses. Finally, we acknowledge the enduring support for the construction and operation of the LHC, the CMS detector, and the supporting computing infrastructure provided by the following funding agencies: SC (Armenia), BMFWF and FWF (Austria); FNRS and FWO (Belgium); CNPq, CAPES, FAPERJ, FAPERGS, and FAPESP (Brazil); MES and BNSF (Bulgaria); CERN; CAS, MoST, and NSFC (China); MINCIENCIAS (Colombia); MSES and CSF (Croatia); RIF (Cyprus); SENESCYT (Ecuador); ERC PRG and PSG, TARISTU24-TK10 and MoER TK202 (Estonia); Academy of Finland, MEC, and HIP (Finland); CEA and CNRS/IN2P3 (France); SRNSF (Georgia); BMFTR, DFG, and HGF (Germany); GSRI (Greece); MATE and NKFIH (Hungary); DAE and DST (India); IPM (Iran); SFI (Ireland); INFN (Italy); MSIT and NRF (Republic of Korea); MES (Latvia); LMTLT (Lithuania); MOE and UM (Malaysia); BUAP, CINVESTAV, CONACYT, LNS, SEP, and UASLP-FAI (Mexico); MOS (Montenegro); MBIE (New Zealand); PAEC (Pakistan); MSHE, NSC, and NAWA (Poland); FCT (Portugal); MESTD (Serbia); MICIU/AEI and PCTI (Spain); MOSTR (Sri Lanka); Swiss Funding Agencies (Switzerland); MST (Taipei); MHESI (Thailand); TUBITAK and TENMAK (T\"{u}rkiye); NASU (Ukraine); STFC (United Kingdom); DOE and NSF (USA).

\begin{sloppypar}
\setlength\emergencystretch{\hsize}
\hyphenation{Rachada-pisek} Individuals have received support from the Marie-Curie programme and the European Research Council and Horizon 2020 Grant, contract Nos.\ 675440, 724704, 752730, 758316, 765710, 824093, 101115353, 101002207, 101001205, and COST Action CA16108 (European Union); the Leventis Foundation; the Alfred P.\ Sloan Foundation; the Alexander von Humboldt Foundation; the Science Committee, project no. 22rl-037 (Armenia); the Fonds pour la Formation \`a la Recherche dans l'Industrie et dans l'Agriculture (FRIA) and Fonds voor Wetenschappelijk Onderzoek contract No. 1228724N (Belgium); the Beijing Municipal Science \& Technology Commission, No. Z191100007219010, the Fundamental Research Funds for the Central Universities, the Ministry of Science and Technology of China under Grant No. 2023YFA1605804, the Natural Science Foundation of China under Grant No. 12535004, and USTC Research Funds of the Double First-Class Initiative No.\ YD2030002017 (China); the Ministry of Education, Youth and Sports (MEYS) of the Czech Republic; the Shota Rustaveli National Science Foundation (Georgia); the Deutsche Forschungsgemeinschaft (DFG), among others, under Germany's Excellence Strategy -- EXC 2121 ``Quantum Universe" -- 390833306, and under project number 400140256 - GRK2497; the Hellenic Foundation for Research and Innovation (HFRI), Project Number 2288 (Greece); the Hungarian Academy of Sciences, the New National Excellence Program - \'UNKP, the NKFIH research grants K 131991, K 138136, K 143460, K 143477, K 147557, K 146913, K 146914, K 147048, TKP2021-NKTA-64, and 2025-1.1.5-NEMZ\_KI-2025-00004, and MATE KKP and KKPCs Research Excellence and Flagship Research Groups grants (Hungary); the Council of Science and Industrial Research, India; ICSC -- National Research Centre for High Performance Computing, Big Data and Quantum Computing, FAIR -- Future Artificial Intelligence Research, and CUP I53D23001070006 (Mission 4 Component 1), funded by the NextGenerationEU program, the Italian Ministry of University and Research (MUR) under Bando PRIN 2022 -- CUP I53C24002390006, PRIN PRIMULA 2022RBYK7T (Italy); the Latvian Council of Science; the Ministry of Science and Higher Education, project no. 2022/WK/14, and the National Science Centre, contracts Opus 2021/41/B/ST2/01369, 2021/43/B/ST2/01552, 2023/49/B/ST2/03273, and the NAWA contract BPN/PPO/2021/1/00011 (Poland); the Funda\c{c}\~ao para a Ci\^encia e a Tecnologia (Portugal); the National Priorities Research Program by Qatar National Research Fund;  MICIU/AEI/10.13039/501100011033, ERDF/EU, ``European Union NextGenerationEU/PRTR", projects PID2022-142604OB-C21, PID2022-139519OB-C21, PID2023-147706NB-I00, PID2023-148896NB-I00, PID2023-146983NB-I00, PID2023-147115NB-I00, PID2023-148418NB-C41, PID2023-148418NB-C42, PID2023-148418NB-C43, PID2023-148418NB-C44, PID2024-158190NB-C22, RYC2021-033305-I, RYC2024-048719-I, CNS2023-144781, CNS2024-154769 and Plan de Ciencia, Tecnolog{\'i}a e Innovaci{\'o}n de Asturias, Spain; the Chulalongkorn Academic into Its 2nd Century Project Advancement Project, the National Science, Research and Innovation Fund program IND\_FF\_68\_369\_2300\_097, and the Program Management Unit for Human Resources \& Institutional Development, Research and Innovation, grant B39G680009 (Thailand); the Eric \& Wendy Schmidt Fund for Strategic Innovation through the CERN Next Generation Triggers project under grant agreement number SIF-2023-004; the Kavli Foundation; the Nvidia Corporation; the SuperMicro Corporation; the Welch Foundation, contract C-1845; and the Weston Havens Foundation (USA).
\end{sloppypar}
\end{acknowledgments}

\bibliography{auto_generated}

\cleardoublepage \appendix\section{The CMS Collaboration \label{app:collab}}\begin{sloppypar}\hyphenpenalty=5000\widowpenalty=500\clubpenalty=5000\cmsinstitute{Yerevan Physics Institute, Yerevan, Armenia}
{\tolerance=6000
A.~Gevorgyan\cmsorcid{0000-0003-2751-9489}, A.~Hayrapetyan, V.~Makarenko\cmsorcid{0000-0002-8406-8605}, A.~Tumasyan\cmsAuthorMark{1}\cmsorcid{0009-0000-0684-6742}
\par}
\cmsinstitute{Marietta Blau Institute for Particle Physics, Vienna, Austria}
{\tolerance=6000
W.~Adam\cmsorcid{0000-0001-9099-4341}, L.~Benato\cmsorcid{0000-0001-5135-7489}, T.~Bergauer\cmsorcid{0000-0002-5786-0293}, M.~Dragicevic\cmsorcid{0000-0003-1967-6783}, S.~Gundacker\cmsorcid{0000-0003-2087-3266}, A.K.~Guven\cmsorcid{0009-0004-5670-5138}, P.S.~Hussain\cmsorcid{0000-0002-4825-5278}, M.~Jeitler\cmsAuthorMark{2}\cmsorcid{0000-0002-5141-9560}, N.~Krammer\cmsorcid{0000-0002-0548-0985}, A.~Li\cmsorcid{0000-0002-4547-116X}, D.~Liko\cmsorcid{0000-0002-3380-473X}, M.~Matthewman, J.~Schieck\cmsAuthorMark{2}\cmsorcid{0000-0002-1058-8093}, R.~Sch\"{o}fbeck\cmsAuthorMark{2}\cmsorcid{0000-0002-2332-8784}, M.~Shooshtari\cmsorcid{0009-0004-8882-4887}, M.~Sonawane\cmsorcid{0000-0003-0510-7010}, N.~Van~Den~Bossche\cmsorcid{0000-0003-2973-4991}, W.~Waltenberger\cmsorcid{0000-0002-6215-7228}, C.E.~Wulz\cmsAuthorMark{2}\cmsorcid{0000-0001-9226-5812}
% ...
\par}
$^{1}$Also at Yerevan State University, Yerevan, Armenia\\
$^{2}$Also at Technische Universit\"{a}t Wien, Vienna, Austria\\
% ...
\end{sloppypar}
%%% END EDITABLE REGION %%%
% skeleton_end
\end{document}